\documentclass[conference, a4paper]{IEEEtran}

\IEEEoverridecommandlockouts                          
\usepackage{graphicx}
\usepackage{multirow}
\usepackage{amsmath,amssymb,latexsym}
\usepackage{rotating}
\usepackage{lettrine}
\usepackage{textcomp}

\usepackage{hyperref}
\usepackage{url}
\usepackage{siunitx}
\usepackage{tikz}
\newcommand*\circled[1]{\tikz[baseline=(char.base)]{
            \node[shape=circle,draw,inner sep=1pt] (char) {#1};}}

\usepackage{xcolor}

\usepackage{lipsum}
\usepackage[pscoord]{eso-pic}
\newcommand{\placetextbox}[3]{
  \setbox0=\hbox{#3}
  \AddToShipoutPictureFG*{
    \put(\LenToUnit{#1\paperwidth},\LenToUnit{#2\paperheight}){\vtop{{\null}\makebox[0pt][c]{#3}}}%
  }%
}%

\title{Seamless Redundancy for High Reliability Wi-Fi}

\author{
    \IEEEauthorblockN{
    Gianluca Cena\IEEEauthorrefmark{1}\href{https://orcid.org/0000-0003-0084-5321}{\includegraphics[scale=0.65]{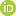}},
    Stefano Scanzio\IEEEauthorrefmark{1}\href{https://orcid.org/0000-0001-7643-2342}{\includegraphics[scale=0.65]{orcid_16x16.png}}, Dave Cavalcanti\IEEEauthorrefmark{2}\href{https://orcid.org/0000-0002-8613-4602}{\includegraphics[scale=0.65]{orcid_16x16.png}}, and Valerio Frascolla\IEEEauthorrefmark{3}\href{https://orcid.org/0000-0002-4256-2955}{\includegraphics[scale=0.65]{orcid_16x16.png}}}
    \IEEEauthorblockA{\IEEEauthorrefmark{1}National Research Council of Italy (CNR--IEIIT), Italy.} 
    \IEEEauthorblockA{\IEEEauthorrefmark{2}Intel Labs, Intel Corporation, Hillsboro, OR, USA.} 
    \IEEEauthorblockA{\IEEEauthorrefmark{3}Intel Labs, Intel Deutschland, Neubiberg, Germany.}
    Email: gianluca.cena@cnr.it, stefano.scanzio@cnr.it, dave.cavalcanti@intel.com, valerio.frascolla@intel.com}

\begin{document}
\placetextbox{0.5}{1}{This is the author's version of an article that has been published in this journal.}
\placetextbox{0.5}{0.985}{Changes were made to this version by the publisher prior to publication.}
\placetextbox{0.5}{0.97}{The final version of record is available at \href{https://doi.org/10.1109/WFCS57264.2023.10144228}{https://doi.org/10.1109/WFCS57264.2023.10144228}}%
\placetextbox{0.5}{0.05}{Copyright (c) 2023 IEEE. Personal use is permitted.}
\placetextbox{0.5}{0.035}{For any other purposes, permission must be obtained from the IEEE by emailing pubs-permissions@ieee.org.}%

\maketitle
\thispagestyle{empty}
\pagestyle{empty}

\begin{abstract}
By removing wire harness, Wi-Fi is becoming increasingly pervasive in every aspect of our lives, in both the consumer and industrial worlds. Besides flexibility, the recent high efficiency and extremely high throughput versions managed to close the performance gap with Ethernet. However, it still lags behind Ethernet for what concerns dependability. To this aim, the ultra high reliability study group has been recently formed.

This paper reports on some preliminary ideas and proposals about the ways seamless redundancy can be exploited to make \mbox{Wi-Fi} more reliable, yet retaining a good degree of backward compatibility with existing network infrastructures.
\end{abstract}


\section{Introduction}
\label{sec:introduction}
A common way to improve dependability of systems and networks is to exploit redundancy
(space, frequency, and time) \cite{PROC19-Pro}.
The High-Availability Seamless Redundancy (HSR) and Parallel Redundancy Protocol (PRP), defined by IEC 62439-3 \cite{IEC62439.3-22},
permit to increase availability and reliability of Ethernet-based industrial networks, also achieving fault-tolerance.
These solutions were taken as the basis for defining redundancy mechanisms for time-sensitive networking (TSN).
In particular, IEEE 802.1CB \cite{IEEE8021CB}, also known as Frame Replication and Elimination for Reliability (FRER),
defines a quite generic framework for redundancy, which can also describe legacy PRP and HSR.
Although FRER scope is not limited to IEEE 802.3 (Ethernet), its optimal usage with other network technologies, like the wireless ones, is not readily apparent.

It must be remarked that the redundancy protocols listed above were conceived for wired networks, where frame losses are typically negligible.
So, they are mainly meant to cope with failures that affect physical links (wires) and intermediate equipment (switches).
Such phenomena are usually \textit{permanent}, which means that human intervention is required to restore the system to the original state, e.g., by replacing broken hardware.
In the meanwhile, the surviving redundant path ensures applications uninterrupted operation, which is essential for mission-critical systems.
Unlike spanning tree (STP and RST), \textit{seamless redundancy} achieves zero-time intervention, 
which makes it suitable for time-critical systems as well.

About one decade ago, the use of seamless redundancy was suggested 
also
for high-performance wireless local area networks (WLAN),
like those based on IEEE 802.11 \cite{IEEE80211-20}.
In \cite{WFCS12-RENT} a proposal was made to employ a pair of \mbox{Wi-Fi} links tuned on different channels as cable replacements, 
each one implemented by means of a wireless client associated to the related access point (AP).
By using conventional PRP equipment (RedBoxes) on the two sides of the connection, seamless redundancy is achieved on air, which permits adoption in safety networks.
In fact, although the black channel does not demand extreme reliability, relatively short outages in \mbox{Wi-Fi} communication may cause timeouts to expire, 
which in turn uselessly trigger the safety-related functions, undermining the overall system availability \cite{ETFA11-CERE}.

Subsequent works postulated that seamless redundancy could be embedded directly in IEEE 802.11 specifications, as proposed for \mbox{Wi-Red} \cite{2016-TII-WiRed}.
By doing so, duplication avoidance mechanisms can be used to lower bandwidth consumption,
which is quite useful in the $\SI{2.4}{GHz}$ band.
In \mbox{Wi-Red}, a redundant STA (RSTA) associates and exchanges frames with a redundant AP (RAP). 
Both the RSTA and the RAP include two (or more) subSTAs (embedded in the same case), tuned on distinct bands/channels, whose behavior closely resembles legacy \mbox{Wi-Fi}.
Frames are duplicated on the transmitting side and deduplicated on the receiving side by a link redundancy entity (LRE),
which coordinates sub-STAs' operation and aborts ongoing transmissions on all channels as soon as an ACK is received for the frame being sent.

More recently, the adoption of TSN concepts in \mbox{Wi-Fi} has been envisaged to make it suitable for time-aware applications as well.
In particular, the use of FRER was considered as a way to support seamless redundancy with no (or minimal) changes to the IEEE 802.11 specification \cite{CSCN21}.
Although \mbox{Wi-Fi 6/6E} is at the moment exploited in new designs for automation systems,
the forthcoming \mbox{Wi-Fi 7}, with its Multi-Link Operation (MLO) \cite{IEEEWCL23}, 
is probably the best option for implementing redundancy in reliable wireless links.

It is essential to remark that seamless redundancy is applied to wired and wireless networks with quite different goals.
In fact, in the latter case it is mainly meant to cope with frame losses due to disturbance,
including both interference caused by nearby wireless devices and electromagnetic noise generated by high-power industrial equipment.
Above phenomena are usually \textit{temporary}, which means that no human intervention is generally required to recover from them.

In the real world, it is quite usual that about $10\div 20\%$ of the transmission attempts performed on air fail, which is clearly unacceptable for any applications.
To make communication reliable, a confirmation mechanism is customarily included in the MAC layer of every wireless protocol.
On frame arrival, the recipient must return an ACK frame.
If the related timeout expires before the ACK is heard, the sender transmits the frame again, until it either succeeds or the retry limit is exceeded.
Leveraging diversity, e.g., the Minstrel algorithm that dynamically selects the best modulation and coding scheme, helps further.
Automatic repeat request (ARQ) mechanisms lower the amount of losses seen by data-link users dramatically, but they also increase transmission latency tangibly, making it unpredictable and often too long for control applications.
Seamless redundancy (along with ARQ) improves both frame losses (it is sufficient that the frame arrives on one channel) and latency (the user is always delivered the fastest copy of every frame) at the same time.

It must be noted that, if fault tolerance has to be ensured to \mbox{Wi-Fi} through PRP, communication equipment must be replicated.
For example, two distinct (R)APs can be deployed, to which two distinct (R)STAs are associated,
as in \cite{WFCS12-RENT}.
In this way, any faults affecting a single device (a problem to the power supply, a broken antenna, etc.) do not disrupt communication.
Distinguishing between reliability and fault tolerance is essential to understand what follows.
The former aims at mitigating temporary disturbance, narrowing the gap between \mbox{Wi-Fi} and Ethernet in terms of the communication quality perceived by applications (e.g., packet loss ratio and deadline miss ratio).
The latter is instead meant to counteract permanent failures, ensuring uninterrupted communication to distributed systems that are expected to be resilient.
In the vast majority of the cases, what is sought from the next generation of \mbox{Wi-Fi} equipment is just a reliable behavior.
This is certainly true for the consumer world, but it also holds in many industrial contexts.
In fact, fault tolerance is expensive \cite{ODVA20-Ditz}, and is only adopted when/where it is really needed.

Neglecting cost, one may argue that fault tolerance implies reliability.
There is, however, another aspect to consider: PRP, HSR, and FRER are static solutions, and redundant paths are typically decided when the network architecture is designed.
This is quite reasonable, since only a careful planning can provide a reasonable certainty about the effectiveness of redundancy, but it somehow betrays the \mbox{plug\&play} nature of Ethernet and \mbox{Wi-Fi}, where prior configuration is not a prerequisite for communication (contrary to TSN).
Typically, all it is needed to connect a node to the network backbone is related to IP (address, netmask, and default gateway)
and security aspects (e.g., service set ID and WPA password).
In a sizeable number of cases, it would be nice to enable reliability through redundancy without requiring the users any additional knowledge about the network.
Besides multimedia communication with mobile devices, also non-safety and non-mission-critical data communication with mobile cobots 
(i.e., not requiring fault tolerance) could benefit from such a feature.

In the following, some preliminary ideas are presented that may help setting the path for the upcoming activities of the IEEE 802.11 Ultra High Reliability (UHR) study group, which will work on the definition of the next WLAN generation (\mbox{Wi-Fi 8}).
Following this introduction, some relevant concepts about redundancy are briefly discussed in Sect.~\ref{sec:red_basics}, 
while the concept of High Reliability STA (HR STA) will be proposed in Sect.~\ref{sec:red_scenario}, where its operation will be also sketched.

\section{Redundant Communication Basics}
\label{sec:red_basics}

Below, two standard solutions aimed at supporting redundancy in IEEE 802 networks are briefly reviewed.

\subsection{Wi-Fi MLO}
One of the key features introduced by 
the IEEE 802.11be task group (\mbox{Wi-Fi 7})
is multi-link operation (MLO).
In particular, a multi-link device (MLD), either STA (non-AP MLD) or AP (AP MLD), 
may send its traffic on two (or more) \textit{affiliated} STAs, also denoted Low MAC (L-MAC), whose behavior is mostly independent 
(unlike channel bonding, where transmission/reception take place exactly at the same time on all the involved channels).
We will focus on simultaneous transmit and receive multi-link multi-radio (STR MLMR) solutions, 
where transmission of different frames by the affiliated STAs (on distinct bands/channels) may overlap partially or totally.

In the current (draft) version of the specification, MLO is mainly intended to increase throughput
and decrease latency,
by letting a single entity in the MLD, the Upper MAC (\mbox{U-MAC}), 
decide at runtime which one among underlying \mbox{L-MACs} must deal 
with the transmission of any buffered frame.
A multi-link element (MLE) is defined, included in Beacon and Association frames, that encodes the information needed to deal with this function.

Using MLOs for sending replicated copies of the frames is in theory possible, 
even though a clear description about the way this can be done is not included in the draft specification.

\subsection{TSN FRER}
Among the many features TSN adds to Ethernet, frame replication and elimination (FRER) aims at improving reliability 
by providing an agreed way to set redundant communication paths between end-points.
The concept of compound stream is introduced, made up of several member streams.
Streams can be split by using frame replication, or joined by using frame elimination.
To enable these operations, frames in a stream are identified by sequence numbers, encoded in a suitable R-TAG.
FRER functions are implemented in both end-nodes and TSN switches.
By suitably configuring them, redundant paths can be set, which provide fault tolerance.

\section{Application Scenarios for Reliable Wi-Fi}
\label{sec:red_scenario}
We consider contexts where infrastructure \mbox{Wi-Fi} networks are connected to a wired Ethernet backbone and possibly to the Internet through routers.
We will assume that at least one of the two communication end-points relies on \mbox{Wi-Fi}.
The case where both communicating devices use \mbox{Wi-Fi} (associated to either the same or distinct APs) is not irrelevant.
A \mbox{Wi-Fi} node that supports seamless redundancy will be denoted for short HR STA or HR AP, depending on its role.
Their behavior resembles RSTA and RAP in \cite{2016-TII-WiRed}, respectively.
To comply with \mbox{Wi-Fi 7}, they are implemented as multi-link devices (MLD), and basically consist of a pair of affiliated STAs.
Every frame for which reliability is demanded is sent by the HR STA/AP on all its affiliated links at the same time.

Similarly to the access category (AC) foreseen by \mbox{Wi-Fi} to specify the frame QoS,
a \textit{reliability category} (RC) must be additionally defined that tells whether or not redundancy is exploited.
Its exact definition is still to be decided: in fact, more-than-duplex redundancy is possible
(also thanks to the new \SI{6}{GHz} band), which means that different degrees of reliability can be envisaged.

The very first aspect to be taken into account when exploiting seamless redundancy to enhance reliability 
is where the frame duplication and deduplication functions have to be located.
Below we will consider three scenarios, characterized by increasing complexity.
What we seek in perspective is a single, unified solution, to be incorporated in \mbox{Wi-Fi 8}.
Only reliable frame exchanges are considered, as  
those that do not need UHR service follow
the default 
\mbox{Wi-Fi} behavior.

\begin{figure}[t]
    \begin{center}
 	\includegraphics[width=1\columnwidth]{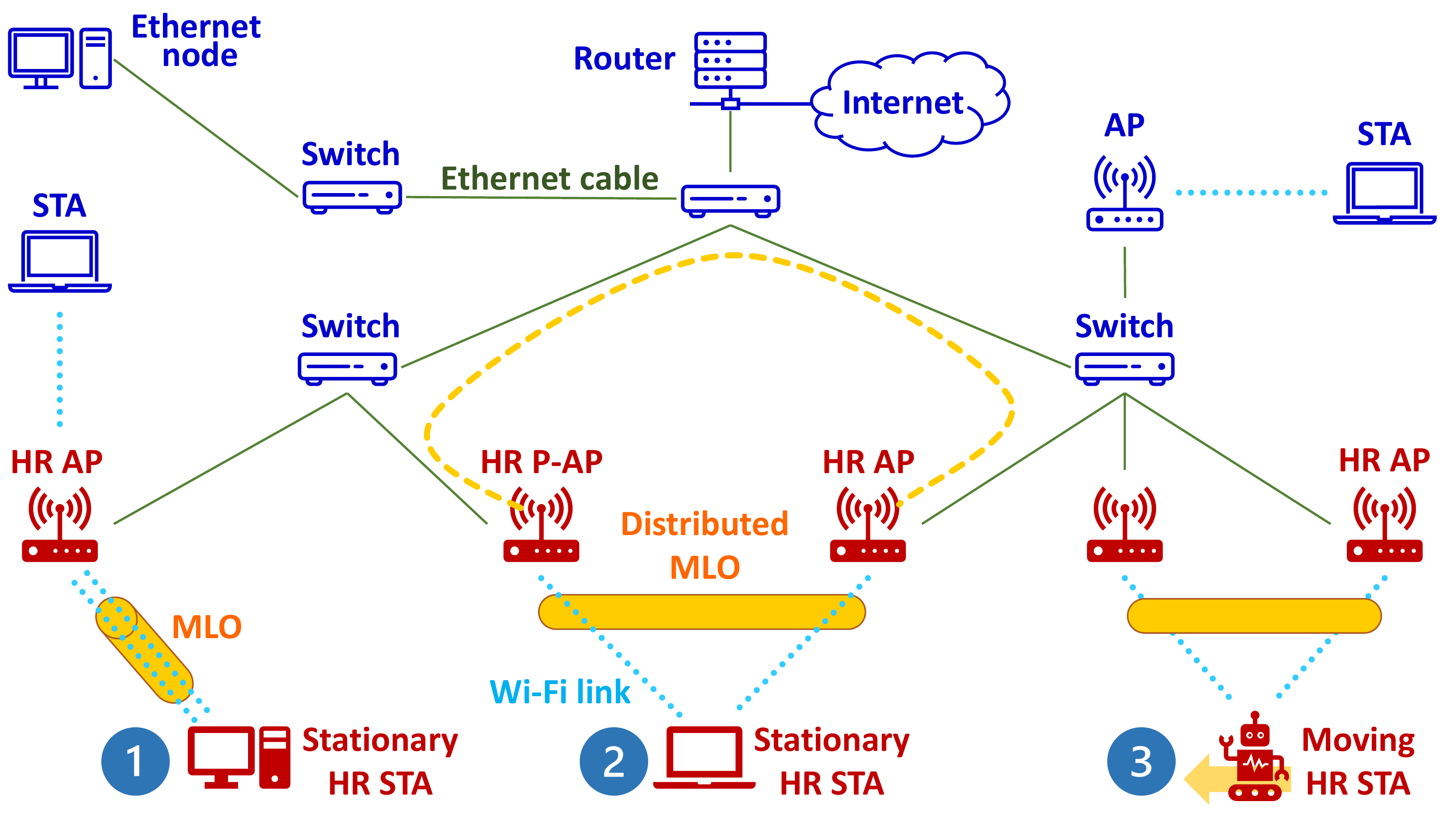}
    \end{center}
    \vspace{-0.3cm}
    \caption{Three relevant scenarios for HR Wi-Fi: 1) Stationary -- Single AP; 2) Stationary -- Multiple AP; and 3) Moving -- Multiple AP.}
    \vspace{-0.2cm}
    \label{fig:arch}
\end{figure}

\subsection{Stationary Wireless Device -- Single AP}

In the first scenario, depicted on the left side of \mbox{Fig.~\ref{fig:arch} \circled{1}},
all the affiliated STAs of an HR STA are associated to the same HR AP.
By exploiting the advertising MLO functions, based on multi-link elements (MLE), 
both end-points of the redundant wireless link are completely aware of what is needed to support seamless redundancy operations.
If required, the MLE format could be augmented to include additional information.

All duplication-deduplication operations are performed 
in the MLD of the
HR STA and the associated HR AP.
This means that any HR STA is seen by the other devices in the network as a conventional node,
and this holds for legacy Ethernet and \mbox{Wi-Fi} devices, as well as for other HR STAs associated to either the same or other HR APs.
They will be unaware of seamless redundancy and just see a more reliable end-to-end path.
Every frame sent by the HR STA is deduplicated by the HR AP before being relayed on Ethernet or to another associated STA.
To make redundancy completely transparent, one of the STAs in the HR STA is selected as the \textit{primary STA}, 
and all frames exiting the associated HR AP include its MAC address as the source address (SA).

When the target is another HR STA associated to the same HR AP,
the latter will duplicate the frame again on the related multi-link.
Similarly, every frame arriving from Ethernet (or from another STA) is duplicated by the HR AP before being sent on the multi-link to which the target HR STA is associated.
The destination address (DA) of incoming Ethernet frames coincides with the MAC address of the primary STA of the HR STA,
but in the relay process on \mbox{Wi-Fi} it is changed to match the MAC address of the actual receivers (affiliated STAs).

Encoding information about RC into the frames sent on air, e.g., as an MLE, could make operation easier.
This is not out of question, as UHR is planned for the next \mbox{Wi-Fi} version, which has still to be defined.
However, this goes beyond the level of detail foreseen for this paper.

\subsection{Stationary Wireless Device -- Multiple AP}

In this second, static scenario, shown in the middle of \mbox{Fig.~\ref{fig:arch} \circled{2}},
the STAs affiliated to the HR STA are associated to different HR APs.
As a consequence, duplication/deduplication can not be performed directly by such APs, as they only see part of the frame copies.
This implies that, unlike the previous case, MLO can not be exploited to support redundancy.
For this reason a ``distributed MLO'' concept has been conceived.

Redundancy as per TSN could be exploited, by leveraging FRER to perform duplication and deduplication in some point of the network located between the two HR APs and the destination node.
To limit bandwidth consumption, FRER should be implemented in the nearest common ancestor of the different HR APs associated to the given HR STA.
To this purpose, an R-TAG is added by the involved HR APs to the frames sent on Ethernet, 
which permits them to suitably travel on the wired portion of the network.
Doing so is not trivial, especially in the case when wireless devices are moving, as it requires that: 
1) the wired portion of the network employs TSN-compliant switches (which is hardly feasible/simple in the brownfield),
and 2) FRER paths are configured in advance for all possible communication end-points.

A more practical solution, which preserves \mbox{plug\&play} behavior, is to have frame replication and elimination performed entirely by a single AP, chosen among those affiliated to the involved HR APs,
that will be termed \textit{primary AP} (P-AP) for the distributed MLO.
A sensible choice is to select as P-AP the AP associated to the primary STA of the HR STA.
To forward the replicated copies to the P-AP for processing, 
a new \textit{relay} tag (Y-TAG) is defined that implements sort of a source routing for redundancy in Ethernet.
Besides a reserved value for the EtherType (on \SI{2}{B}), which permits to detect the \mbox{Y-TAG}, 
it includes the next destination address (NA, on \SI{6}{B}) to which the frame must be forwarded in the subsequent relay,
a sequence number to manage elimination of replicated copies (on \SI{2}{B}), 
plus (possibly) ancillary information, similar to what is needed to deal with MLO redundancy.

When a copy of the frame is received (on air) from the associated HR STA,
all the involved non-primary HR APs relay it to the P-AP over Ethernet using the Y-TAG.
Upon reception of one of these frames, 
the P-AP removes the Y-TAG and reconstructs the original frame by using NA as the new DA.
It has all the information needed to understand that this is a copy that employs distributed MLO.
Thus, deduplication can be performed in the same way as if the frame were received on air from the HR STA associated to the P-AP.
If the frame is not discarded, it is relayed to destination.

\begin{figure}[t]
    \begin{center}
 	\includegraphics[width=1\columnwidth]{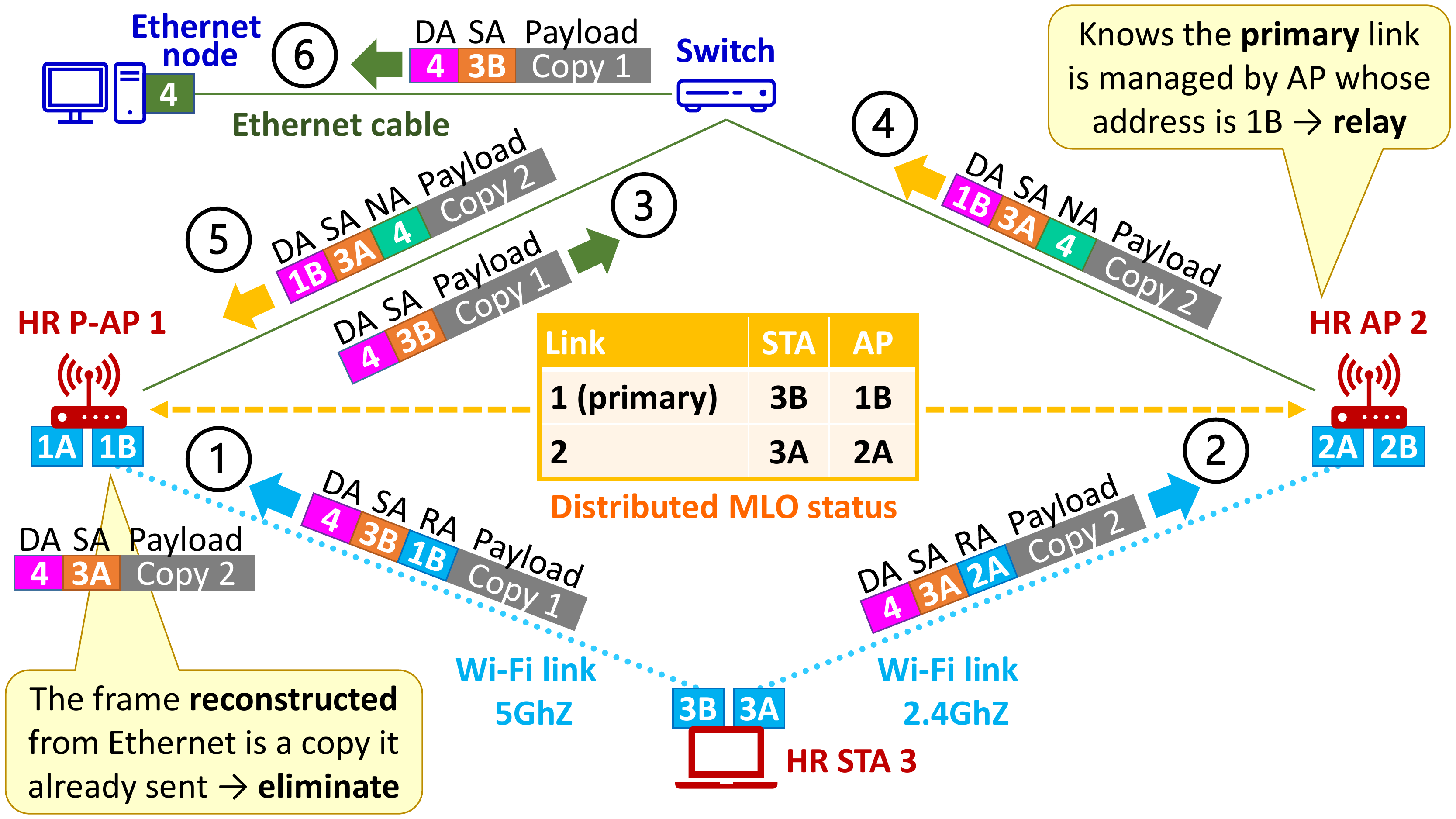}
    \end{center}
    \vspace{-0.3cm}
    \caption{Sample HR distributed multi-link operation: all copies are relayed to the primary AP (using the Y-TAG), which performs elimination.}
    \vspace{-0.2cm}
    \label{fig:distmlo}
\end{figure}

An example of distributed MLO is sketched in Fig.~\ref{fig:distmlo}.
HR STA 3 is associated to two distinct HR APs (1 and 2).
A frame for which reliable service is demanded is sent
contextually by both its affiliated STAs (3A and 3B).
The copy on the left (from 3B) is sent on air to the P-AP (1B).
When arriving at the non-primary AP (2A), the copy on the right (from 3A) is relayed to the P-AP (1B) on Ethernet.
By exploiting the Y-TAG, where NA points to the final destination (4), the original frame can be reconstructed.
The copy that arrives first to the P-AP (the one on the left, in the example) is relayed to destination, the other is discarded.

The reverse direction (from an Ethernet node to an HR STA associated to multiple HR APs) is managed similarly, with the P-AP that relays copies of the frame to the other involved HR APs for parallel transmission (on air) to the target HR STA.

\subsection{Moving Wireless Device -- Multiple AP}

The third scenario, on the right side of \mbox{Fig.~\ref{fig:arch} \circled{3}}, closely resembles the second one, but moving nodes are now present.
Ensuring reliable interconnection on air to roaming devices makes the link transition between different states of the former two kinds, 
and implies that associations to APs change over the time.
So that communication quality is not impaired, reassociation must be performed one link at a time.
In other words, when this process starts on one of the STAs of the HR STA, 
the other affiliated STAs are not allowed to do so until the ongoing reassociation finishes.
Eventually, all STAs will be associated to a new HR AP.

Generally speaking, three types of transitions may occur while roaming: 
the former turns a single AP scenario into a multiple AP one (MLO to distributed MLO), while the second performs the reverse operation (MLO is managed more effectively than distributed MLO, and hence the HR STA should do its best to return to the single AP scenario).
A third type of transition
occurs when switching between different multiple AP scenarios.
Overall, the protocol must deal with such transitions smoothly and flawlessly.
To this aim, the use of artificial intelligence to select the best time/way for reassociation is likely to bring tangible benefits.
Instead, single AP to single AP transitions are not possible, as reassociation is performed one link at a time.

\section{Conclusion}
High reliability is the next Holy Grail for \mbox{Wi-Fi}, as witnessed by the formation of the UHR study group.
The solution we propose focuses on enhancing the Wi-Fi operation
without requiring any changes to the already deployed Ethernet networks.
Conversely, frame replication is performed only on the wireless portion of the paths,
where losses due to disturbance actually occur.
We deem that this is advantageous for non-critical systems, where fault tolerance is not needed.

A new kind of \mbox{Wi-Fi} devices is envisaged, denoted HR STAs and HR APs, each one including two (or more) affiliated STAs.
Qualitative analysis was performed for three relevant scenarios.
When the HR STA is associated to a single HR AP, MLO can be used to support seamless redundancy.
On the contrary, when it is associated to multiple HR APs, a mechanism denoted distributed MLO has been introduced that adheres to the \mbox{plug\&play} philosophy of Ethernet/\mbox{Wi-Fi} and makes the involved APs transparently manage duplication and deduplication, without requiring any prior configuration.
The case where HR STAs are allowed to move relies on the previous two mechanisms, 
by defining reliable reassociation procedures.
Our proposal is clearly in a very early stage, and requires a formal definition, a careful investigation about achievable performance, and some proof of correctness.

\bibliographystyle{IEEEtran}
\bibliography{bibliography}

\end{document}